\documentclass{ws-ijmpd}

\begin{document}

\markboth{Tammi \& Hovatta}
{Thermal flares in AGNs and microquasars}

%
\catchline{}{}{}{}{}
%


\title{DISK-JET CONNECTION IN AGNS AND MICROQUASARS: \\
THE POSSIBILITY OF THERMAL FLARES IN THE CENTER}

\author{JONI TAMMI\footnote{Email: joni.tammi@iki.fi} \; \& 
  TALVIKKI HOVATTA\footnote{Current address: 
  Department of Physics, Purdue University, 525 Northwestern Avenue, 
  West Lafayette, IN 47907, USA. Email: thovatta@purdue.edu}}
\address{Aalto University Mets\"ahovi Radio Observatory, \\ 
  Mets\"ahovintie 114, FI-02540 Kylm\"al\"a, Finland}


\maketitle
\begin{history}
  \received{10 December 2009}
  \revised{1 February 2010}
\end{history}


\begin{abstract}
  We discuss the possibility of thermal flares in centers of AGNs and
  microquasars. We present preliminary results of an ongoing study
  trying to assess the feasibility of a hypothesis suggesting that
  certain flares observed in these sources originate in the very
  centers of the systems and not in the relativistic jets. Using a
  simple toy model we reproduce optical flares with lightcurves very
  similar to those observed in the sources. The model suits especially
  well those cases where only the latter peak of a double-peaked
  optical flare has a radio counterpart.
\end{abstract}



\section{Introduction} \label{sec:intro}

Many characteristics of flaring in active galactic nuclei (AGN) and
microquasars are best explained with relativistic motion of the
radiating regions in jets. Especially for microquasars the variability
is often best explained by variations in the structure of the
accretion disk and the jets; as discussed by many authors throughout
this volume, the connection between the disk and the large-scale jets
is well established even if the details are not yet fully known.

Our starting point for this report is the general model proposed for
both the microquasars and AGNs (see
refs.~\refcite{MirabelRodriguez1998,Marscher2002}, and references
therein). The connection between the accretion disk and the radio jet
in AGNs has been discussed from different angles, \emph{e.g.}, in
refs.~\refcite{Marscher2002}--\refcite{Chat08}, and in general the
scenario is believed to work along the following lines: part of the
accretion disk breaks off, and while the inner hot X-ray emitting
matter falls beyond the event horizon (causing a decrease in the X-ray
flux (see ref.~\refcite{Marscher2002} and references therein) the
outer parts get injected into the jet that is collimated and
accelerated before becoming visible at what is called the core of the
jet. There the jet plasma is believed to be compressed and heated by a
standing shock wave,\cite{Marscher2008,Chat09} causing a
multifrequency flare and a new knot becoming visible in the
jet.\cite{Savol02,Chat08} Apart from the jet core this is the same
process that was proposed for the microquasar GRS 1915+105 already
earlier.\cite{MirabelRodriguez1998} In this report we suggest an
additional component to the scenario.

Our study was motivated by the observations and idea of Miller-Jones
et al.,\cite{TwinPeaks} who found that certain double-peaked flares in
the Cygnus X-3 are best explained by treating the first flare as a
product of an outbreak of disk wind, later followed by a second flare
when a new jet element becomes visible further away in the jet. Here we study
the possibility that this model is at work also in AGNs and that -- in
addition to getting swallowed by the black hole or being injected into
the jet -- part of the collapsing accretion disk matter erupts from
the center as a wind-like outflow. We test if this outburst of matter --
originally having temperatures comparable to the inner parts of the
disk -- could produce observable effects or, in the most radical case,
cause a flare comparable to those happening in the jet, thus leading
to double-peaked flare. Modeling the causes of such
an event is beyond the scope of this early-phase report -- at this
point we focus on estimating what observational signatures a sudden
outflow of matter in the center of an AGN or microquasar could
produce.

In the following we describe our hypothesis for the first flare in the
center (Sec.~\ref{sec:thermal}) and discuss the expected signatures
and, in Sec.~\ref{sec:doubleflare}, discuss possible predictions of
the model concentrating on the application to AGNs. An example is
given in Sec.~\ref{sec:example} using BL Lacertae as a test case.


\section{Toy model for a thermal outburst in the center} \label{sec:thermal}

In this early phase of the study we use a very simplified model to see
if a thermal flare with reasonable parameters can, even in principle,
cause any of the observed features. For this test we take a
``spherical cow'' approach: we describe the radiating plasma as a
homogeneous sphere expanding at a constant speed and compute the
blackbody and the bremsstrahlung emission (taking the free-free
absorption and the light-travel time effects into account) at
different frequencies as a function of time.

For simplicity we take the matter to consist only of electrons and
positrons initially having a temperature comparable to that of the
accretion disk ($T_0\sim 10^{4-6}$ K for AGNs\cite{BonningEtAl}), and
calculate the matter density from the initial radius of the sphere
($r_S < R_0 \lesssim 3 r_{\rm S}$, where $r_{\rm S} = 2 GM/c^2$ is the
Schwartzschild radius of the black hole) and the mass of the
collapsing disk, estimated to be of the order of 1--10 $M_\odot$,
assuming the disk to contain mass accreted since the last major burst
--typically occurring once every couple of
years\cite{HovattaEtAl2007}-- at a rate of a few $M_\odot$/yr.

In the present model the sphere of plasma immediately starts to expand
with a constant speed --which we assume to be mildly relativistic, up
to 0.5 c (see, \textit{e.g.}, ref.~\refcite{Tombesi})-- and also cool
down due to adiabatic losses.


\section{Double flares -- observational features and predictions} \label{sec:doubleflare}

\paragraph{Different flare characteristics.} 

As the mechanisms and the environments related to the first flare in
the center and the second flare at the core of the jet are different,
also the flares should differ from each other from spectral as well as
temporal points of view. Firstly, with typical AGN parameters the
sphere is optically thin for optical photons from the beginning,
making the flare bright in the optical waveband almost
immediately. Millimeter and radio flux, on the other hand, peak weeks
to months later and reach flux levels of only small fractions of those
of the optical peak. It is also important to notice that the
timescales or radiative signatures of first flare would not be
affected by similar Lorentz boosts or Doppler shifts than the flares
in the jet.

\paragraph{Single or double flares, depending on frequency.} 

Whereas a thermal flare would most likely be observed and interpreted
as an optical flare without radio counterpart, a later flare (at the
radio core) is always expected to be bright throughout the
spectrum. This means that depending on the observing frequency, the
two flares would be observed either as a single- or double-peaked
flares (at the radio and optical wavelenghts, correspondingly).

\paragraph{Location of the core and delay between the flares.}

The delay between the two flares, $\Delta t$, would depend on the
distance of the jet core from the central black hole, on the speed
(and acceleration) of the jet, and on the light-travel-time
effects. Assuming, for simplicity, instant acceleration of the jet to
the speed corresponding to the apparent velocity $v_{\rm app}$ at
which the knots in the post-core jet (making an angle $\theta$ with
the line of sight) are seen to travel, the separation of the core is
simply $r_{\rm core} = \Delta t v_{\rm app} / \sin \theta$. Taking
into account the finite time required for the acceleration of the jet,
this may overestimate the actual distance.

\paragraph{Dip in the degree of polarization.}

Because the radiation from the first thermal flare is taken to be
unpolarized, and if the intensity of polarized emission is not
changed, the rise of the unpolarized radiation leads to a
corresponding decrease of the polarization. As a zeroth-order estimate
one can estimate the dip in the polarization simply from \( P = P_0 /
( 1 + I_{\rm flare}) \), where $P_0$ and $I_{\rm flare} = \Delta I /
I_0$ are the pre-flare polarization and the brightness increase
compared to the pre-flare level.

\paragraph{Flux levels start to rise before the second flare.}

If the first flare loads the jet with matter, this could lead to
the flux starting to rise already before the second flare. On one hand
this is due to the particles in the plasma being energized by the
strong magnetic turbulence near the center providing promising
conditions for efficient second-order Fermi acceleration,\cite{TD09}
leading to increased synchrotron emissivity. On the other hand, the
acceleration of the ''proto jet'' emitting region as a whole leads to
increased Lorentz boosting of the emitted radiation.

\paragraph{New ''knot'' in the jet after the second flare.}

Finally, when the jet-injected matter reaches the location of the VLBI
core of the radio jet, the standing shock (assumed to lie in the core)
compresses the plasma and accelerates particles on short timescales,
causing a multifrequency flare. After passing through the core, the
radiating plasma continues to travel along the jet and is now seen
as a bright knot.\cite{Savol02,Chat08} From this point on the ejected
plasma can be described using various shock-in-jet models.

\section{An example: BL Lacertae} \label{sec:example}

\begin{figure}[tH]
\centerline{\psfig{file=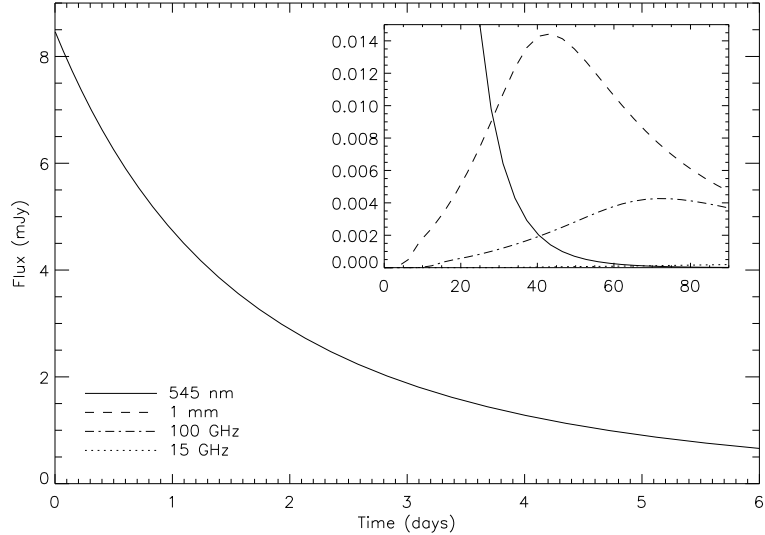,width=0.8\linewidth}}
\vspace*{8pt}
\caption{Example lightcurves from a thermal flare in BL Lacertae for
  optical (solid line) and three radio frequencies as given in the
  plot. The dimmer and slower radio flares are only visible in the
  small panel showing the micro-Jansky level peaks months after the
  optical mJy-level flare. See text for details.\label{f1}}
\end{figure}

Let us illustrate the points raised in Sec.~\ref{sec:doubleflare} in
the case of the blazar BL Lac ($z=0.069$, $M \approx 10^8$ $M_\odot$).
By assuming the outbursting mass to be $M_{\rm in} = 10\, M_\odot$
(the major outbursts in the source happening on average every 2--3
years,\cite{NieppolaEtAl2009} leading to required mass accretion rate,
3--5 $M_\odot$/yr), the expansion speed $v_{\rm exp} = 0.4$ c, initial
radius $r_0 = 1.5\, r_{\rm S}$, initial temperature $T_0 = 10^5$ K and
adiabatic index $\gamma = 4/3$, we get an optical flare rising very
fast (within hours) to 8.5 mJy and declining in a few days with $S
\propto e^{-t/2}$, approximately, as shown in Fig.~\ref{f1}. At the
radio frequencies, however, the source becomes optically thin much
later and has a much lower peak flux density keeping the radio
signature of the flare invisible under the strong and variable
nonthermal flux from the jet. In essence this would be seen as an
optical flare without a radio counterpart.

To have even an order-of-magnitude reality check before detailed data
analysis, we compare this to observations of BL Lac made with with the
Kanata telescope in October
2008.\footnote{http://f.hatena.ne.jp/kanataobslog/20090106201125,
  cited 1 Dec 2009.}  Beginning on 11th of October 2008, we see an
optical flare rising from 8.4 mJy to 17.0 mJy within a few days, the
flux increasing by approximately 8.6 mJy. This is remarkably close to
the 8.5 mJy obtained from our zeroth-order model using first-guess
input parameters. For the duration of the flare the polarization
degree decreased from 18 \% to 11 \%. Using the same initial
polarization, we obtain the value of 9 \%, still interestingly near
the observed one. The comparison is, of course, very crude, but it is
sufficient to keep the proposed scenario plausible.


\section{Discussion and Conclusions} \label{sec:discussion}

In addition to the presented case for BL Lac, we have tested our toy
model for different AGNs and microquasars. The results (in
preparation) are promising: even with our very crude spherical-cow
model we have been able to reproduce lightcurves similar to those
observed both in AGN and microquasar objects. Some sources, however,
seem to be completely beyond the scope of the flux levels and
timescales obtainable with the present simple model. We are currently
collecting multifrequency data for detailed testing and further
development.

Further developments of the model will enable testing the X-ray
emission (enabling comparison with the observed anticorrelation
between the X-ray and radio fluxes \cite{Chat09}). Also the possible
role of nonthermal radiation, even in the case of a mostly thermal
flare, needs to be studied. Furthermore, the physicality of the model
will be improved by including different geometries and dynamics for
the outflow, as well as taking into account the presence of,
\emph{e.g.}, the dust torus for AGNs and companion star in
microquasar, as well as the differences in the density and composition
of the environment and the matter in the coronae of the black hole and
the accretion disk.

Finally we comment on two recent reports that are relevant to the
points presented in Sec.~\ref{sec:doubleflare}. Firstly, Marscher et
al.\cite{Marscher2008} reported an example of an optical double peak
with a single radio burst followed by a jet element in the jet of BL
Lacertae. The first flare was shown to happen in the acceleration and
collimation region before the radio core of the jet, following an
explosive event near the black hole; they explain the first optical
flare as a product of the increased radiation from the accelerating
plasma blob before the core. In our model the flare would take place
closer to the black hole -- otherwise our model assumes everything to
happen as described by Marscher et al.\cite{Marscher2008} The
published data are, however, still too sparse to either support or
disprove our model; further continuous and dense multifrequency
observations are needed to distinguish the possible non-jet flares
from those happening in the outflows or in the accretion
disk. Secondly, a very recent article by Villforth et
al.\cite{VillforthEtAl2009} published polarization data of the BL Lac
object OJ287. Their lightcurves included many peaks consisting mostly
of unpolarized flares. In this source the variability is considered to
be linked to the dynamics of a binary black hole with the secondary
black hole disrupting the accretion disk (see their paper for a review
of different models suggested for the source). Their results could
provide interesting and very useful tool in testing the proposed
model, as they emphasize the repeated double-peaked outbursts with the
first peak having no radio counterpart. Furthermore, they report dips
in the optical polarization during the burst. We acknowledge, however,
that the toy model presented here does not seem to be able to
reproduce the brightness of OJ287 flares. We dare not yet speculate
whether improved modeling of the black hole's environment could help,
or can this kind of an approach apply even in theory in an object
suspected to harbor a binary supermassive black hole
(ref.~\refcite{VillforthEtAl2009} and references therein).

To conclude, based on the preliminary results and within the limits of
currently available data we cannot rule out the possibility that in
some microquasars and AGNs certain flares can be due
to a thermal or partly thermal flare associated with the explosive
event that also dismisses parts of the accretion disk and injects
material into the jets. However, more data and improvements for the
model are needed to satisfyingly estimate the feasibility of
non-jet flares in these sources.

\section*{Acknowledgments}
We thank the Kanata telescope observation blog for their openness in
providing example data. Furthermore, Dr.~Tuomas Savolainen is
acknowledged for pointing out the possibility for using dip of the
polarization degree in testing the model.



\end{document}